# The in-flight spectroscopic performance of the *Swift* XRT CCD camera

J. P. Osborne[a], A. P. Beardmore[a], O. Godet[a], A. F. Abbey[a], M. R. Goad[a], K. L. Page[a], A. A. Wells[a], L. Angelini[b], D. N. Burrows[c], S. Campana[d], G. Chincarini[d], O. Citterio[d], G. Cusumano[e], P. Giommi[f], J. E Hill[b,g], J. Kennea[c], V. LaParola[e], V. Mangano[e], T. Mineo[e], A. Moretti[d], J.A. Nousek[c], C. Pagani[c], M. Perri[f], P. Romano[d], G. Tagliaferri[d], F. Tamburelli[f]

[a] University of Leicester, University Road, Leicester, LE1 7RH, UK
[b] NASA-GSFC, Greenbelt, MD 20771, USA
[c] Pennsylania State University, 525 Davey Lab, University Park, PA 16802, USA
[d] INAF-Osservatorio di Brera, Via Bianchi 46, 23807 Merate, LC, Italy
[e] INAF-IASF, Via U. La Malfa 153, 90146 Palermo, Italy
[f] ASI-ASDC, Via G. Galilei, I-00044 Frascati, Italy
[g] USRA, 10211 Wincopin Circle, Suite 500, Columbia, MD 21044-3432, USA

**ABSTRACT**

**The Swift X-ray Telescope (XRT) focal plane camera is a front-illuminated MOS CCD, providing a spectral response kernel of 144 eV FWHM at 6.5 keV. We describe the CCD calibration program based on celestial and on-board calibration sources, relevant in-flight experiences, and developments in the CCD response model. We illustrate how the revised response model describes the calibration sources well. Loss of temperature control motivated a laboratory program to re-optimize the CCD substrate voltage, we describe the small changes in the CCD response that would result from use of a substrate voltage of 6V.**

## 1. INTRODUCTION

The Swift gamma-ray burst satellite[1] was successfully launched from Cape Canaveral on 2004 November 20. It carries a sensitive coded mask Burst Alert Telescope[2] to locate new GRBs in the 15-150 keV range. When a burst from an unknown source is detected and a slew to its location is possible, Swift autonomously re-points to bring the GRB within the field of view of the two narrow field instruments: the X-ray Telescope[3,4] and the UV/Optical Telescope[5]. Observations with these instruments may start within about a minute of the burst. Swift thus uniquely provides routine prompt observations of GRBs and their afterglows, it also automatically transmits information from all three instruments via the TDRSS satellites and the Gamma-ray burst Coordinate Network[6] to observers and robotic telescopes around the world.

The Swift X-ray Telescope consists of a thermally controlled and optically monitored carbon fibre telescope tube, an X-ray mirror system consisting of 12 concentric gold-coated shells of 3.5 m focal length in a Wolter-1 configuration, and a Focal Plane Camera Assembly (FPCA) housing a cooled e2v CCD-22 having 600x600 image pixels located behind an optical blocking filter with an optical transmission of 0.25%. The FPCA includes a door, and autonomous sun shutter, calibration sources and a substantial mass of Al proton shielding (which also reduces thermal variations). The CCD is mounted on a thermoelectric cooler connected via a heat pipe to an external radiator, a system designed to have an operating temperature of -100°C. Analogue and digital electronics process the CCD signals. The XRT is able to autonomously select a spectral data collection mode according to the current count rate so as to avoid the effects of pile-up[7]. At the highest count rates Photodiode (PD) mode reads out the entire CCD with no spatial information at 0.14 ms resolution; at intermediate count rates Windowed Timing (WT) mode preserves one spatial dimension to give a time resolution of 1.8 ms; at lower count rates Photon Counting (PC) mode gives full spatial information at 2.5 s resolution.

High quality spectral information is provided in all of these modes, although the limited spatial information in PD and WT modes requires the use of different event pattern libraries to characterize the charge distribution over neighbouring pixels in the three modes[4]. The XRT is sensitive over 0.2-10 keV, having ~135 cm$^2$ effective area at 1.5 keV. It has a 23.6 arcmin$^2$ field of view and a point spread function half energy width of 15.4-18 arcsec over 0.5-4.0 keV. CCDs of an identical design are in use in the EPIC MOS cameras on ESA's XMM-Newton satellite.

The XRT has measured the early X-ray light curves and spectra of all of the GRB afterglows at which it has been promptly pointed, some 32 so far. These observations have revealed previously unsuspected multiple breaks[8,9] and flares[10] in the early X-ray light curves, as well as absorption in excess of that due to our own Galaxy, consistent with sites of star formation as expected for the collapsar model of the long GRBs[11]. The XRT has also provided the first accurate location of a short GRB, suggestively near an elliptical galaxy, raising the possibility that these events are due to binary neutron star spiral-in and collision[12].

In this paper we present the XRT CCD in-flight spectral calibration and describe improvements to the response model made available as response matrices through the HEASARC caldb[13].

## 2. PRE-FLIGHT CALIBRATION

The XRT flight CCD underwent an extensive spectral calibration at the Leicester calibration facility. High statistical quality spectra were obtained using lines from carbon (0.27 keV) to copper (8.04 keV). In addition, limited spectral calibration was achieved at the MPE Panter facility using the XRT electronics and with the mirror in place. The spectra were used to steer the tuning of physical models of the interactions of X-rays in the e2v CCD-22. These models are used by our Monte-Carlo response matrix generator to make mode- and event pattern-specific response matrices. The physical model includes the effects of: incident photon transmission through the CCD electrode structure, energy-dependent absorption, fluorescent emission, variations in charge cloud spreading and recombination in the different device layers, surface charge loss, charge transfer efficiency, and collection of the charge into pixels accounting for the event characterization thresholds[14].

Before launch our response model was used to make response matrices for all three spectroscopic modes and for various event pattern (grade) selections (PC mode grades 0, 0-4 and 0-12, WT mode grades 0 and 0-2, PD mode grade 0, 0-2 and 0-5). Our model was highly successful at describing the pre-launch calibration spectra.

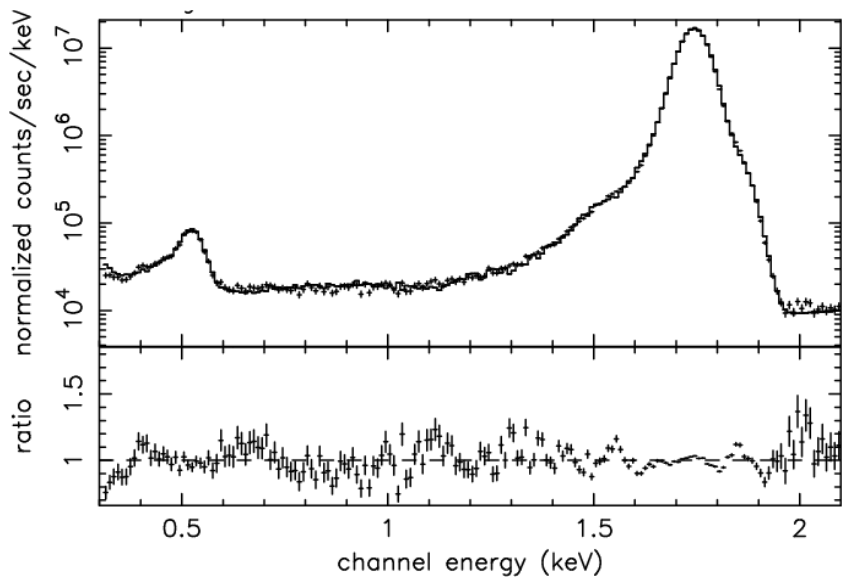

Figure 1. XRT PC mode single pixel event spectrum measured at Leicester from the Si Kα and Si Kβ lines (1.74 and 1.84 keV) with background O Kα (0.525 keV) and Bremsstrahlung continuum, together with pre-flight modeled response.

# 3. IN-FLIGHT EXPERIENCES

Before the XRT reached its operational state, and before the CCD was cooled, the XRT thermo-electric cooler (TEC) power supply system apparently failed. All attempts to revive the TEC have been unsuccessful, and the XRT has to rely on passive cooling via the heat pipe and radiator in combination with enhanced management of the spacecraft orientation to reduce the radiator view of the sunlit earth. The design CCD operating temperature was -100°C. In flight CCD temperatures of -75 to -45°C have been seen. Elevated CCD temperatures cause increased thermal charge which is manifested as excess counts at very low energy, increased output from hot pixels, and visibility of areas of raised dark current (due to oxide damage during ground calibration) which would be invisible at the design operating temperature. In principle the spectral response can also be degraded; but the effects are barely discernable, and were reduced by a tuning of the on-board bias map algorithm in a flight software upload on 2005 June 13. In addition there is a clear dependence of the gain on the temperature, as illustrated below. We find that CCD temperatures below -47°C result in good quality data, and we find no measurable temperature dependence of the spectral response width.

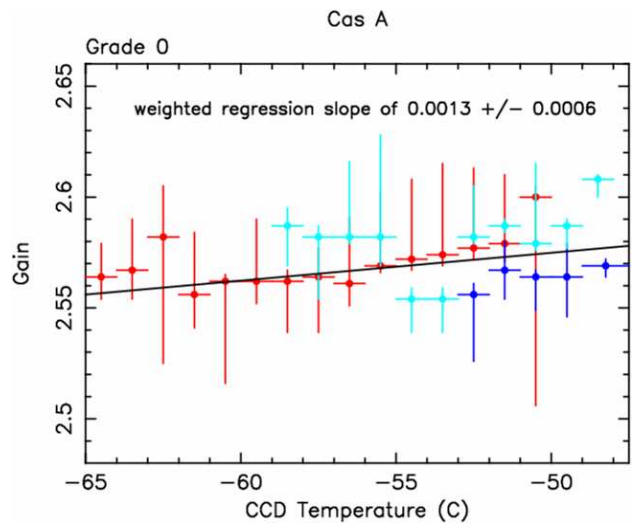

Figure 2. The XRT gain as a function of temperature measured in PC mode using emission lines of Si, S, Ar, Ca and Fe from the supernova remnant Cas A.

It was also quickly realized that an additional source of diffuse background radiation was present in the XRT PC mode images. In severe cases this appears as an off-centre broad ring with a differently centred blob together covering a significant fraction of the field of view. X-ray-like events were seen below 0.5 keV (and predominately below 0.2 keV) at the start and end of the target visibility windows. These times correspond to spacecraft pointing directions close to the sunlit earth, and it appears that significant optical light is reaching the CCD when pointing within ~120 degrees of the bright earth. Optical photons provide charge in addition to the thermal noise, and can alter the effective energy scale of the CCD. This effect is excluded in scientific data reduction by the avoidance of times of low bright earth angle.

At 05:22 UT on 2005 May 27 the count rate in the XRT increased suddenly causing a switch from < 1 c/s in PC mode to saturation in piled-up PD mode. XRT images at this time show substantial and uneven light illumination. Subsequent PC mode images show new bright pixels, one bright column, and two bright column segments (the precise effects are temperature dependant). Charge leakage from the top of the bright column affects its immediate neighbours. This event has been diagnosed as due to a particle (micrometeoroid) scattering off the mirror system to hit several CCD pixels. The XRT internal light source shows that there is no substantive damage to the optical filter. There have been similar events observed in the XMM EPIC MOS CCDs. The bright pixels and columns have been vetoed on-board, although this is not possible for PD mode, which at present is not in use. This event had no apparent impact on the spectroscopic performance of the XRT, or on its ability to image and locate new GRBs.

## 4. IN-FLIGHT CALIBRATIONS

Spectroscopic calibrations after launch have been performed with the XRT FPCA door source (Fe55) prior to the door opening on 2004 Dec 11, and a number of well known celestial objects, many of which are calibration sources for other X-ray observatories. The astronomical calibrators used are listed below. Calibration observations were made primarily before 2005 April 5, although routine calibration observations after this date continue to use around 5% of the time.

| **Object** | **Type** | **Mode** | **Purpose** | **Exposure (ks)** |
|---|---|---|---|---|
| RXJ 0720.4-3125 | Pulsar | PC | Low energy response | 19.6 |
| RXJ 1856.5-3754 | Neutron Star | PC | Low energy response | 5.4 |
| PKS 0745-19 | Cluster of galaxies | PC | Low energy shelf and effective area | 29.7 |
| 2E 0102-7217 | SNR | PC<br>WT<br>LrPD | Effective area and gain | 24.3<br>25.4<br>6.0 |
| Cas A | SNR | PC<br>WT | Effective area and gain | 5.8<br>45.7 |
| 3C 273 | Quasar | WT | Effective area and cross-calibration | 10.0 |
| Crab | SNR | WT<br>LrPD | Effective area and gain | 5.8<br>25.0 |
| PSR 0540-69 | Pulsar | PC | Effective area | 33.9 |

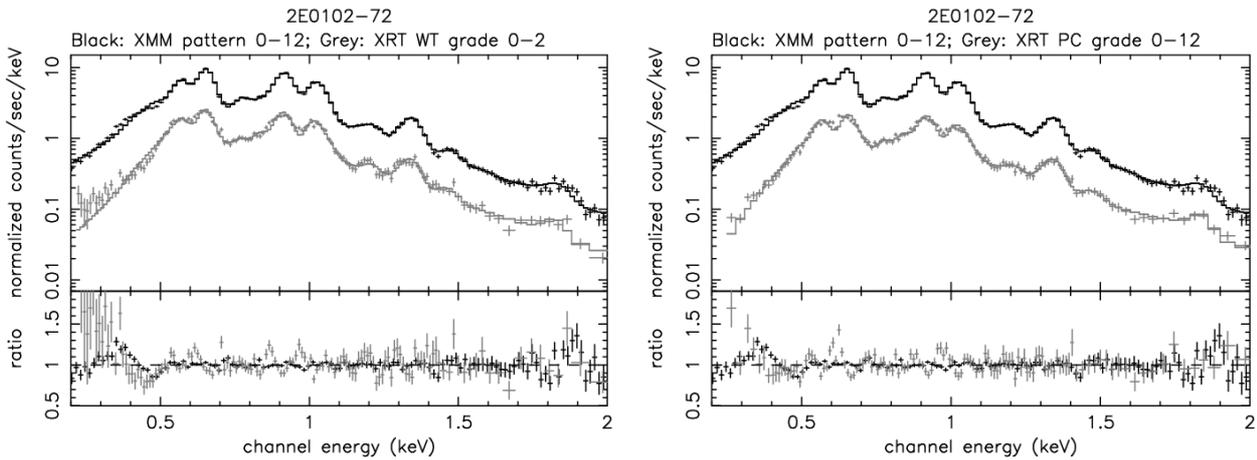

Figure 3. a) *Left:* XMM MOS and Swift XRT WT grade 0-2 spectra of the line-rich SNR 2E0102-72 fit with current response matrices (XRT v007). b) *Right*: As left, but for Swift XRT PC mode grades 0-12 (1-4 pixel events).

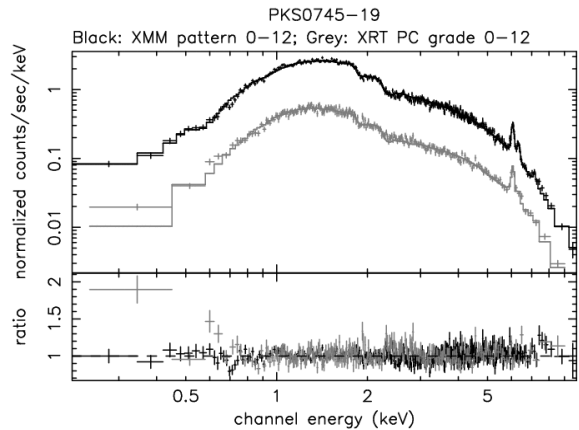

Figure 4. XMM MOS and Swift XRT PC grade 0-12 spectra of the galaxy cluster PKS0745-19 fit with current response matrices (XRT: v007, with an effective area file still to be released).

Fits to some of the celestial calibration sources using the first post-launch response matrices (v007) are shown in Figs 3 and 4. The line-rich supernova remnant 2E 0102-72 shows residuals below 20% above 0.4 keV in WT and PC modes, at lower energies the need for modifications to the response shelf is apparent. The absorbed cluster of galaxies PKS 0745-19 ($N_H$=4.2x$10^{21}$ cm$^{-2}$) has residuals below 20% up to 7 keV. This source was observed to calibrate the shelf of the response; response model improvement is clearly needed here. The low energy end of the response is demonstrated by the soft isolated neutron star RXJ 0720-31; the result of using an improved loss function can be seen in Fig 5.

## 5. POST-LAUNCH RESPONSE MODEL DEVELOPMENTS

A typical X-ray spectrum obtained with a CCD for mono-energetic radiation differs significantly from a simple Gaussian, which is expected in the ideal case. It consists of four components: a Gaussian peak, a shoulder on the low energy side of the peak, a shelf extending to low energies, and at the very lowest energies a noise peak (of which only the high energy side may be seen above threshold). The peak, shoulder and shelf are visible in Fig 1.

The shoulder may have two different origins depending on the line energy. At low energies (typically below 1.5 keV in our case), the shoulder is produced by the charge loss at the interface between the $SiO_2$ layer and the active silicon volume of the open electrode. This effect also produces a low energy shelf in the response. Above 1.5 keV, sub-threshold losses become dominant. When a photon interacts within the CCD, after spreading in the volume of the detector, it generates charge which is collected in the depletion region. The charge cloud may spread into more than one pixel. To make the response matrices we stack simulated spectra of monochromatic X-rays. We need to use the pattern recognition process used in the analysis software, for PC mode this is a 3x3 pixel matrix centred on the highest pixel. To avoid noise being included in the charge summation (and to avoid excessive telemetry usage) a threshold is set on-board below which pixels are not considered. Charge below this low energy threshold is lost, thus causing the extension of the peak to lower energies in the shoulder observed[3,4,7,14].

The pre-flight spectral response did not fully reproduce the shelf in general, or the shoulder for X-rays above 1.5 keV for the larger events. Changes in our spectral response code have been made as described below.

**5.1 X-ray absorption in the substrate**

X-ray absorption in the substrate was not included in the Monte-Carlo code when the pre-flight spectral response was made. The fraction of photons interacting in this region is around 30% and 40% respectively at 8 keV and 9 keV. The photons interacting near the interface between the substrate and field-free regions produce a charge cloud; however only a small fraction of this charge is collected, that remaining is lost to recombination and trapping in the substrate region. These interactions make multi-pixel events suffering substantial charge loss; they have been included in the v007 response code and matrices, adopting the procedure of Short et al.[15]

## 5.2 The low energy shelf

The response shelf results from the charge losses at the surface of the open electrode. These losses are a result of an inversion of the potential near the surface due to the charge-state of the oxide surface[15]. No satisfactory physical model exists; nevertheless it is possible to take into account these losses by introducing a loss function which depends on the photon energy and depth of absorption. Although a loss function was used in the v007 response model, we have since used a function based on that given by Eq. 4 in Pavlov and Nousek[17]. Adjustments to the free parameters of this function significantly improved the low energy spectral response, as shown in Fig. 5

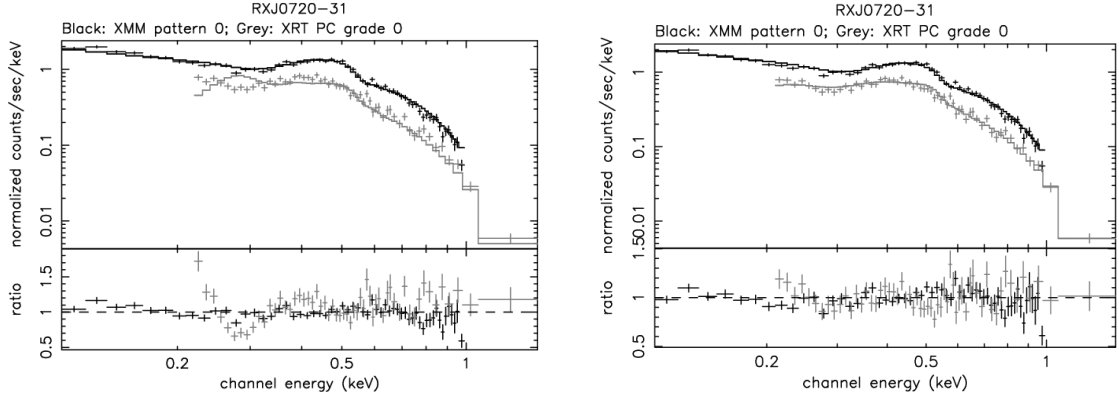

Figure 5. a) Left: XMM MOS and Swift XRT PC grade 0 spectra of the isolated neutron star RX J0720-31 fit with current response matrices (XRT v007). b) Right: As a), but the XRT data fit with a response matrix using an improved surface charge loss function.

## 5.3 The shoulder for high energy X-rays

In the v007 response we adopted a temporary method to reproduce the shoulder by artificially increasing the event split threshold. Fig. 6a shows the resulting fits to the Fe55 door source spectra. The shoulder is not reproduced well.

We have now enhanced the description of the shoulder by modifying the shape of the charge cloud in the field-free region. This had been approximated by a 2-D Gaussian, but the cloud shape could be significantly non-Gaussian here[17]. Fig. 6b shows the good agreement of the model with the door source spectrum when fit using the new cloud shape.

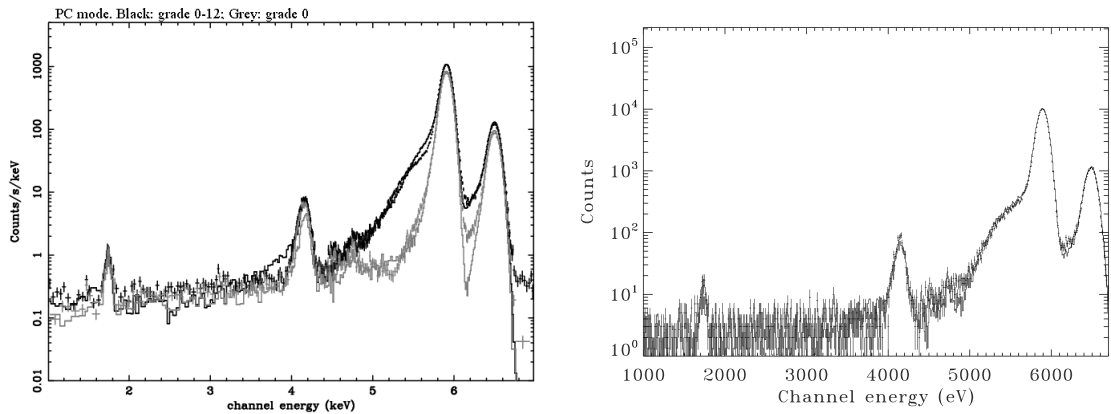

Figure 6. a) Left: The XRT door Fe55 source spectrum in PC mode for single (grade 0, grey) and 1-4 pixel events (grade 0-12, black) compared with the v007 response model, the fit is poor for the large events in the response shoulder. b) Right: The 1-4 pixel PC mode door source spectrum much better fit with a new response model using a revised field-free charge cloud shape.

## 6. PLANNED NOISE REDUCTION BY A SUBSTRATE VOLTAGE CHANGE

The higher than anticipated XRT operating temperatures caused by the loss of active cooling cause significant thermally induced noise. This noise can be manifested as low energy events, and thus has the potential to use up valuable on-board storage and telemetry capacity. A reduction in thermal noise can save resources, allow the use of lower energy X-ray events, and reduce any broadening of the spectral response kernel. Such a noise reduction can be achieved by the use of a raised CCD substrate voltage, $V_{ss}$, which can fill traps in interface states (due to inversion) and by reducing the volume of silicon in which carriers are generated.

A laboratory program at the Space Research Centre at the University of Leicester was initiated to investigate the effects on CCD performance of various substrate voltages over a range of temperatures. The XRT flight spare CCD (identical to the flight CCD) was used. A substrate voltage of 0 V has been used in flight since launch. Substrate voltages between 0 V and 8 V were evaluated at 1 V intervals with the CCD at temperatures of -45, -50, -60 and -70°C. Spectra were obtained in both PC and PD modes using a Ti source emitting 4.5 and 4.9 keV Kα and Kβ X-rays. Subsequent work also made use of Al and Cu X-ray sources.

The gain of the CCD reduced approximately linearly with $V_{ss}$ at 0.7% per volt, and charge transfer inefficiency effects were insignificant in the voltage range tested. These effects play no role in the voltage optimization. For the PC mode, increasing substrate voltage caused the noise peak charge to decrease monotonically at all but the highest temperatures, and at all but the lowest temperatures caused the CCD response kernel FWHM to decrease. For the PD mode, increasing substrate voltage caused the noise peak again to fall monotonically, as did the response kernel FWHM; however the width of the noise peak was at a minimum between 5 and 6 V for all temperatures. This work led to the conclusion that a change of substrate voltage from 0 V to 6 V would be optimal, allowing the CCD to be operated at least 5°C warmer for the same dark current.

Increasing the substrate voltage decreases the depletion depth of the CCD; this increases the ratio of multi-pixel to single events. Because X-ray event charge in adjacent pixels can be below threshold, sub-threshold energy loss manifests itself as a low energy shoulder to the main response peak. Higher voltages increase the fraction of multi-pixel events and thus the sub-threshold energy loss, effectively reducing the quantum efficiency; although all of these effects are modest. For events of 1-4 pixels quantum efficiency is reduced to 0.87 of the 0 V value at 8 V measured over the Ti spectral peak, but is only reduced to 0.93 when the shoulder is included.

Using the observed event size distribution for Al, Ti and Cu X-rays to estimate a depletion depth change from 27μ at $V_{ss}$=0 V to 22μ at $V_{ss}$=6 V, new response matrices were made with our Monte-Carlo code. These were used to investigate the effect of an increased substrate voltage on the detected X-ray spectrum of celestial sources. Fig 6 shows simulated spectra of the Crab supernova remnant for the current substrate voltage of $V_{ss}$=0 V, and for $V_{ss}$=6 V; the loss of quantum efficiency can be seen to be very minor. At the time of writing the use of a revised substrate voltage on-board has yet to be tested.

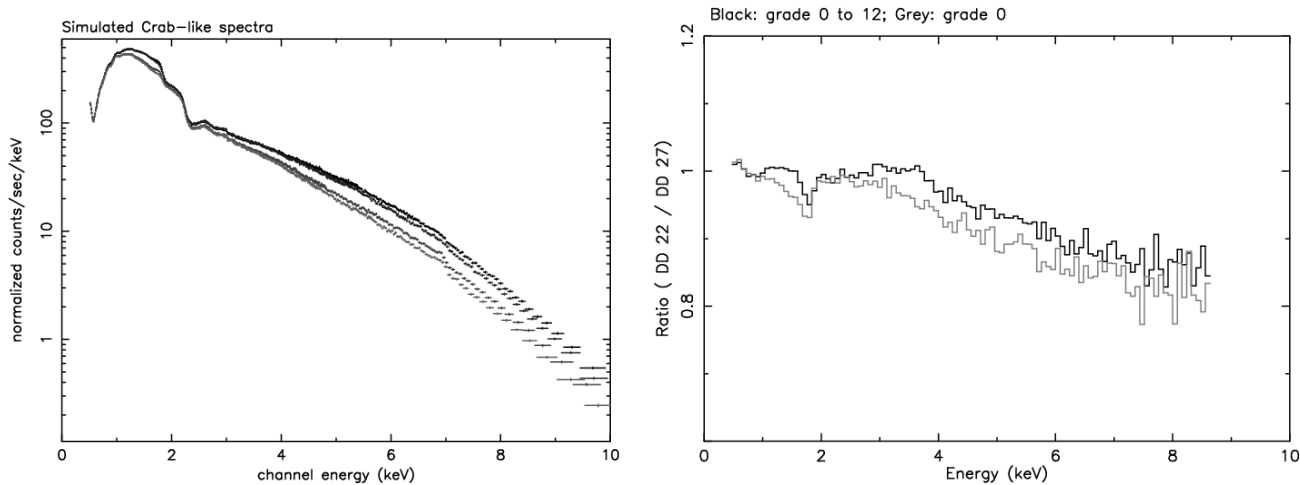

Figure 7. a) *Left:* XRT spectra simulated for a Crab spectral model. From top to bottom: Vss=0 1-4 pixel events, Vss=6 1-4 pixel events, Vss=0 single pixel events, Vss=6 single pixel events. b) *Right:* Ratio of effective area with Vss=6 over area with Vss=0 (top: 1-4 pixel events, bottom: single pixel events).

## 7. CONCLUSIONS

The Swift XRT is scientifically productive and operating efficiently, in spite of elevated noise and an apparent micrometeoroid hit. The pre-flight spectroscopic calibration of the XRT has been shown to be sufficient for initial astrophysical analysis, having been confirmed by flight spectra. Response model developments continue, and improvements to the description of the low energy shelf and response shoulder are described. The effects of the high CCD temperature can be reduced with minor loss of effective area by raising the CCD substrate voltage from 0 to 6 V.